\documentclass[acmsmall]{acmart}

\usepackage{natbib}

\newcommand{\methods}{\textcolor{black}{88.24\%\xspace}}
\newcommand{\por}{\textcolor{black}{56.93\%\xspace}}
\newcommand{\at}{\textcolor{black}{12.67 seconds\xspace}}
\newcommand{\androtestApps}{\textcolor{black}{56\xspace}}
\newcommand{\porWoPhil}{\textcolor{black}{23.24\%\xspace}}

\newcommand{\rwaAT}{\textcolor{black}{278.9 seconds\xspace}}
\newcommand{\rwaMR}{\textcolor{black}{62.03\%\xspace}}
\newcommand{\rwaRR}{\textcolor{black}{54.80\%\xspace}}

\newcommand{\guardianrr}{\textcolor{black}{34\%\xspace}}
\newcommand{\aperr}{\textcolor{black}{11.12\%\xspace}}
\newcommand{\goalexplorerrr}{\textcolor{black}{4.75\%\xspace}}

\newcommand{\flowdroidmr}{\textcolor{black}{58.81\%\xspace}}
\newcommand{\droidreachmr}{\textcolor{black}{9.48\%\xspace}}
\newcommand{\droidreachat}{\textcolor{black}{23.46 seconds\xspace}}
\newcommand{\flowdroidat}{\textcolor{black}{35.06 seconds\xspace}}

\newcommand{\quotes}[1]{``#1''} 

\usepackage[most]{tcolorbox}

\usepackage[acronym]{glossaries}
\glsdisablehyper
\newacronym{OS}{OS}{Operating System}
\newacronym{GUI}{GUI}{Graphical User Interface}
\newacronym{LLM}{LLM}{Large Language Model}
\newacronym{CG}{CG}{Call Graph}
\newacronym{CGs}{CGs}{Call Graphs}
\newacronym{ICC}{ICC}{Inter-Component Communication}
\newacronym{CHA}{CHA}{Class Hierarchy Analysis}
\makeglossaries{}

\usepackage{xcolor}
\usepackage{hyperref}

\usepackage{graphicx}

\usepackage{caption}

\usepackage{balance}

\usepackage{supertabular}

\usepackage{datetime2}

\usepackage{xspace}

\usepackage{listings}
\usepackage{float}
\definecolor{codegreen}{rgb}{0,0.6,0}
\definecolor{codegray}{rgb}{0.5,0.5,0.5}
\definecolor{codepurple}{rgb}{0.58,0,0.82}
\definecolor{backcolour}{rgb}{0.95,0.95,0.92}
\lstdefinestyle{mystyle}{
    backgroundcolor=\color{backcolour},   
    commentstyle=\color{codegreen},
    keywordstyle=\color{magenta},
    numberstyle=\tiny\color{codegray},
    stringstyle=\color{codepurple},
    basicstyle=\ttfamily\footnotesize,
    breakatwhitespace=false,         
    breaklines=true,                 
    captionpos=b,                    
    keepspaces=true,                 
    numbers=left,                    
    numbersep=5pt,                  
    showspaces=false,                
    showstringspaces=false,
    showtabs=false,                  
    tabsize=2
}

\lstdefinelanguage{json}{
    string=[s]{"}{"},
    stringstyle=\color{SchoolColor},
    comment=[l]{:},
    commentstyle=\color{black},
    basicstyle=\normalfont\ttfamily,
    numbers=left,
    numberstyle=\scriptsize,
    stepnumber=1,
    numbersep=8pt,
    showstringspaces=false,
    breaklines=true,
    frame=lines
}
\usepackage[linesnumbered,ruled,lined]{algorithm2e}
\SetKwFor{For}{for (}{)}{}

\usepackage{booktabs}
\usepackage{tikz}
\usepackage{listofitems}

\usepackage{comment}

\usepackage{algorithmic}
\usepackage{graphicx}
\usepackage{textcomp}
\usepackage{xcolor}
\def\BibTeX{{\rm B\kern-.05em{\sc i\kern-.025em b}\kern-.08em
    T\kern-.1667em\lower.7ex\hbox{E}\kern-.125emX}}
\begin{document}

\title{GAPS: Targeted Execution of Android Apps via Static Path Reconstruction}

\author{Samuele Doria}
\email{sdoria@math.unipd.it}
\affiliation{%
  \institution{University of Padua}
  \city{Padova}
  \country{Italy}
}

\author{Alexander Pilgun}
\email{aleksandr.pilgun@uni.lu}
\affiliation{%
  \institution{University of Luxembourg}
  \city{Luxembourg}
  \country{Luxembourg}
}

\author{Jordan Samhi}
\email{jordan.samhi@uni.lu}
\affiliation{%
  \institution{University of Luxembourg}
  \city{Luxembourg}
  \country{Luxembourg}
}

\author{Jacques Klein}
\email{jacques.klein@uni.lu}
\affiliation{%
  \institution{University of Luxembourg}
  \city{Luxembourg}
  \country{Luxembourg}
}

\author{Eleonora Losiouk}
\email{eleonora.losiouk@unipd.it}
\affiliation{%
  \institution{University of Padua}
  \city{Padova}
  \country{Italy}
}

\begin{abstract}

  Dynamically executing target methods in Android applications remains a critical and unresolved challenge. Despite notable advancements in GUI testing, current tools are insufficient for reliably driving execution toward specific target methods.

  To address this challenge, we present GAPS (Graph-based Automated Path Synthesizer), the first system that leverages static, method-guided call graph reconstruction to guide the dynamic, interaction-driven execution of an Android app. GAPS performs a lightweight backward traversal of the call graph, guided by data-flow analysis, to reconstruct paths reaching the target methods. GAPS translates these paths into instructions that guide the execution of the app.

  On the AndroTest benchmark, GAPS statically identifies paths towards \methods\ of the target methods, averaging just \at\ per app, and reaching \por\ of them through dynamic analysis. This exceeds the state-of-the-art tools' one: the model-based GUI tester APE reaches only \aperr, the hybrid tool GoalExplorer reaches \goalexplorerrr, and the LLM-based Guardian reaches \guardianrr. Finally, we applied GAPS to the 50 most downloaded apps from the Google Play Store, achieving an average static analysis time of \rwaAT\ to reconstruct paths towards \rwaMR\ of the target methods and reaching \rwaRR\ of them through dynamic analysis.

\end{abstract}

\maketitle

\section{Introduction}
\label{introduction}

Android dynamic analysis is widely used to study app behavior at runtime, supporting tasks such as vulnerability validation, malware detection, and performance assessment.
By executing an app on a real device or emulator, it can observe behaviors that static inspection may miss.
Its effectiveness, however, depends on exercising the relevant behavior, which is difficult on Android because app behavior is often triggered by \gls{GUI} interactions, lifecycle callbacks, and framework-driven events.
Consequently, dynamic analysis commonly relies on automated exploration to steer execution toward the behavior or code region of interest.

This work focuses on \emph{target methods}: specific methods whose runtime execution must be verified.
Such methods arise in security, testing, and program analysis, e.g., as potentially vulnerable methods reported by static analysis, sensitive API wrappers, or routines whose reachability must be confirmed.
We study the \emph{reachability of target methods}: given an app and a target method, determine whether the method is reachable from a valid entry point and, if so, synthesize and execute interactions that trigger it.

This task is distinct from the goals of existing Android interaction tools. Coverage-oriented GUI testers~\cite{wang2025llmdroid, stoat, collins2021deep} maximize explored screens, events, or statements, while task-oriented agents~\cite{guardian, autodroid, gptdroid, visiontasker} complete user-level goals such as \quotes{add an event to the calendar app}. Neither goal requires executing a preselected method: a tool may obtain high coverage or complete a task while missing the target. In contrast, target method reachability may require reasoning about callbacks, \gls{ICC}, path conditions, data dependencies, and entry points that are not visible from the current screen. Thus, establishing whether a target method can be reached remains an open problem~\cite{browserprivacy, monkeylimitations, androtest}.

Prior Android GUI testing tools illustrate this mismatch. Pseudo-random tools such as Monkey~\cite{monkey} support shallow crash discovery but lack models of app semantics or UI structure. Model-based and hybrid systems, including APE~\cite{ape} and GoalExplorer~\cite{goalexplorer}, improve coverage and bug discovery using adaptive heuristics or statically constructed screen-transition models~\cite{wang2021infrastructure, su2021benchmarking, akinotcho2024mobile}. More recent techniques use machine learning~\cite{ares} or Large Language Models (LLMs), with Guardian~\cite{guardian} achieving strong results on semantic task-completion benchmarks. Yet these systems still optimize for broad exploration or user-task completion, rather than the systematic triggering of a target method~\cite{columbus, androtest, stochtesting, sapienz, timemachine, dynodroid}.

Static analysis is complementary, but existing call-graph and path-reconstruction tools are also insufficient. Whole-program call graphs are costly and often imprecise for Android because of callbacks, lifecycles, reflection, and \gls{ICC}~\cite{cg_soundness, mudflow}. Moreover, a static path does not, by itself, identify the concrete UI actions or runtime conditions required to execute the target method. This motivates a hybrid approach that starts from the target method, reconstructs only relevant paths, and uses them to guide dynamic execution.

We first evaluate whether current state-of-the-art tools already solve this problem. Using the \androtestApps\ open-source apps from the AndroTest dataset~\cite{androtest}, a standard benchmark still used by current Android interaction approaches~\cite{columbus, androtest, stochtesting, sapienz, timemachine, dynodroid, akinotcho2024mobile}, we randomly selected 50 target methods per app and measured both static path reconstruction and dynamic execution. On the static side, FlowDroid~\cite{flowdroid}, the best-performing call graph generator~\cite{cg_soundness}, generated paths for \flowdroidmr\ of the target methods, while DroidReach~\cite{droidreach}, the only available static path-reconstruction tool, generated paths for \droidreachmr. These tools rely on whole-program abstractions with scalability bottlenecks~\cite{mudflow} and fail to extract execution-critical data, such as actionable GUI widget identifiers and runtime branch conditions, that are needed to dynamically trigger paths. On the dynamic side, across three 30-minute runs, Guardian~\cite{guardian} executed only \guardianrr\ of the target methods on average, while APE~\cite{ape} and GoalExplorer~\cite{goalexplorer} reached \aperr\ and \goalexplorerrr, respectively. Thus, current static and dynamic methods are insufficient for target-driven reachability.

To address this gap, we design \textbf{GAPS} (\textbf{G}raph-based \textbf{A}utomated \textbf{P}ath \textbf{S}ynthesizer), a framework that uses static analysis to guide dynamic execution to a target method. Rather than performing whole-program analysis, GAPS constructs a target-oriented, context-sensitive backward call graph only on the code relevant to the target. It applies \emph{data-flow} analysis, including \emph{points-to-analysis} and \emph{constant propagation}, to reconstruct partial paths, resolve conditions, identify triggering UI elements, and translate the results into high-level interaction instructions. GAPS then executes these instructions with its dynamic component, which includes \textbf{PHIL} (\textbf{P}redictive \textbf{H}andler for \textbf{I}nterface \textbf{L}imitations), an integrated agentic reasoning module that handles interface limitations during guided interaction.

On AndroTest, GAPS generates at least one path for \methods\ of the target methods and automatically traverses it, reaching \por\ at runtime. Its static analysis completes in \at\ on average, and dynamic execution never exceeded 5 minutes of runtime. In a real-world security scenario, we used SPECK~\cite{speck} to identify potential vulnerabilities in the 50 most-downloaded Google Play apps, then assessed their reachability with GAPS. It generated paths for \rwaMR\ of the target methods in \rwaAT\ on average and dynamically reached \rwaRR\ of them.

The contributions of the paper are as follows:
\begin{itemize}
    \item We design GAPS, a hybrid framework that performs target-oriented backward static analysis and translates the recovered paths into actionable dynamic interaction steps. We release the framework, evaluation data, and setup in open source: \url{https://anonymous.4open.science/r/GAPS/README.md}.
    \item We evaluate GAPS on AndroTest, the state-of-the-art dataset for benchmarking automated interactions on Android apps. GAPS generates at least one path towards \methods\ of the target methods in \at\ on average, and dynamically reaches \por\ of them without ever exceeding a runtime of 5 minutes.
    \item We demonstrate GAPS in a real-world security use case by measuring the reachability of potentially vulnerable methods across the 50 most-downloaded apps on the Google Play Store. With an average static analysis time of \rwaAT, we statically compute paths for \rwaMR\ of the target methods and dynamically reach \rwaRR.
\end{itemize}

\section{Related Work}
\label{related_works}

Prior work uses the terms \quotes{goal} and \quotes{target} to mean different objectives, including broad coverage, user-task completion, activity or API reachability, vulnerability discovery, or taint propagation. We therefore organize related work by the objective that drives the analysis: static reachability analysis, GUI exploration, user-goal-driven UI automation, and hybrid static-dynamic systems. GAPS addresses a narrower problem than these systems: proving and validating, first statically and then dynamically, whether a selected target method is reachable. It achieves its purpose through an initial static phase that recovers target-specific entry points, path conditions, and triggering UI elements, followed by a dynamic phase that validates whether those paths execute the target method at runtime.

\subsection{Static Reachability Analysis}
\label{cg}

Static analysis approaches for Android apps map callbacks, calls, data dependencies, and \gls{ICC} links to determine whether code locations may be invoked. The well-known FlowDroid~\cite{flowdroid} and Amandroid~\cite{amandroid} build inter-procedural, flow- and context-sensitive abstractions primarily for security and taint tracking; IccTA~\cite{iccta} extends FlowDroid with \gls{ICC} modeling. These whole-program analysis approaches are foundational but resource-expensive on modern apps~\cite{mudflow}. Their outputs remain disconnected from GUI elements and branch-triggering conditions, and they do not indicate whether a path corresponds to a reachable runtime state.

BackDroid~\cite{backdroid} performs backward inter-procedural analysis, but its objective is taint analysis rather than synthesizing and validating an executable path to a target method. It also does not translate static findings into GUI events or runtime checks, and a recent study reports that it performs worse than FlowDroid~\cite{cg_soundness}.

DroidReach~\cite{droidreach} is the closest to our static path-reconstruction approach. It recovers paths from complete call graphs generated by Androguard~\cite{androguard_cg} or FlowDroid, but remains purely static: it does not identify actionable entry points and GUI triggers, infer runtime path conditions, or dynamically validate feasibility. In Section~\ref{rq1}, GAPS produced paths toward \methods\ of the target methods faster than both FlowDroid and DroidReach, while also producing the information needed to drive execution.

\subsection{GUI Exploration}

Automated GUI testers execute UI and system events to expose runtime errors and increase coverage. Random generators such as Monkey~\cite{monkey} require little configuration, while SmartMonkey~\cite{smartmonkey} and Dynodroid~\cite{dynodroid} add heuristics to reduce unproductive inputs. Model-based and learning-based tools, including APE~\cite{ape}, ARES~\cite{ares}, and Q-Testing~\cite{qtesting}, further prioritize the exploration of unseen screens, transitions, or crashes. Even work described as \emph{targeted}~\cite{azim2013targeted} focuses on the exploration strategy rather than on the reachability of target methods.

These tools are effective for broad exploration, but they select actions based on the current GUI state and coverage-oriented heuristics. They do not reconstruct a path to a code-level target or reason about hidden callbacks, \gls{ICC}-induced control flow, guarded branches, or UI states outside the visible screen. We use APE as the representative GUI exploration baseline due to its strong performance in recent comparisons~\cite{wang2021infrastructure, su2021benchmarking, akinotcho2024mobile}. In our evaluation (Section~\ref{rq2}), APE executed only \aperr\ of the target methods, whereas GAPS reached \por\ of them.

\subsection{User-Goal-Driven UI Automation}
\label{llm_ui}

LLM-based UI automation translates natural-language tasks into GUI actions using interface context. GPTDroid~\cite{gptdroid} uses view hierarchies and interaction memory; Autodroid~\cite{autodroid} first collects context through random exploration; VisionTasker~\cite{visiontasker} identifies elements from screenshots; and Guardian~\cite{guardian} adds replanning and state restoration to recover from invalid actions.

These systems are goal-directed at the user-task level, not at the method-reachability level. Their plans are anchored in visible UI semantics and user instructions, so a task may succeed without executing a particular target method, especially when the method is hidden behind callbacks, non-UI components, library code, or guarded branches. Guardian is the strongest dynamic baseline in this family because its replanning can sustain deeper exploration. However, it achieved only \guardianrr\ of the target methods on average across three runs (Section~\ref{rq2}), whereas GAPS achieved \por\ by grounding execution in statically recovered paths.

\subsection{Hybrid Static-Dynamic Systems}
\label{hybrid}

Hybrid Android systems combine static and dynamic analysis approaches, but static information often serves detection or coverage rather than target method reachability. Malware-analysis frameworks such as Andrubis~\cite{andrubis} and AndroPyTool~\cite{andropytool} extract static features and collect sandbox behavior, without converting static results into a path-sensitive runtime plan. Evodroid~\cite{evodroid} uses a static Control Graph Model to guide evolutionary GUI exercising, but its objective is coverage, and its MoDisco-based analysis~\cite{modisco} depends on source-code artifacts. GAPS instead operates on APK and DEX inputs and reaches a target method.

GoalExplorer~\cite{goalexplorer} is the closest hybrid, goal-directed system. It builds a screen-transition graph and guides execution toward targets such as activities, API calls, or program statements. Its guidance is therefore centered on screens and activity transitions. GAPS starts from a method-level target, constructs a demand-driven backward slice of relevant call paths, resolves path conditions, and derives executable UI steps from that path. This distinction is reflected in Section~\ref{rq2}, where GoalExplorer reached only \goalexplorerrr\ of the target methods.
\section{GAPS Design}
\label{design}

GAPS enables the automated generation of interaction instructions that guide an app to invoke a target method. An interaction instruction guides the app through a sequence of UI elements and entry points without requiring any human intervention. GAPS is designed to address common challenges in Android app analysis, such as \gls{ICC}, condition resolution, and GUI event retrieval, working directly on the APK, without requiring access to the source code.

\begin{figure}[t]
    \centering
    \includegraphics[width=\linewidth]{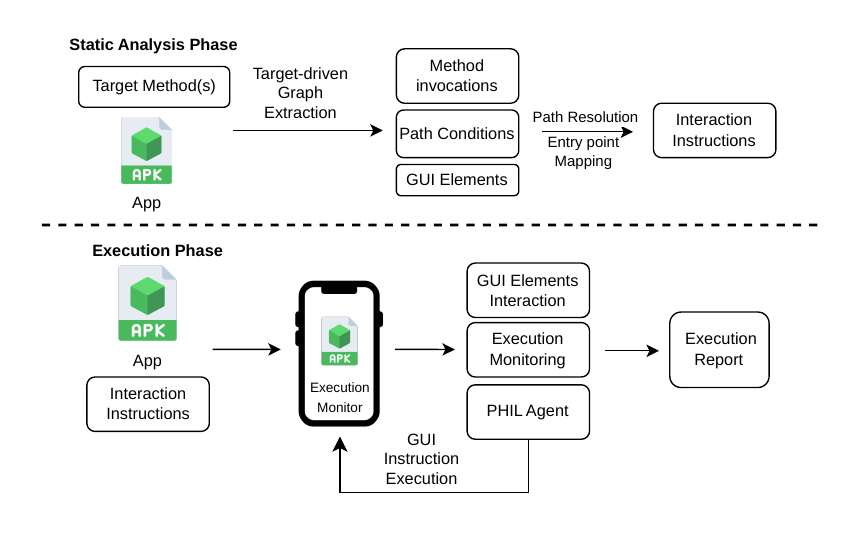}
    \caption{The GAPS execution pipeline. The architecture is divided into two stages: the Static Analysis Phase, which extracts method invocations, conditions, and GUI triggers to generate interaction instructions; and the Execution Phase, which executes the instructions on the app and concludes with an execution report.}
    \label{fig:gaps-overview}
\end{figure}

Figure~\ref{fig:gaps-overview} illustrates GAPS' workflow that consists of two phases: \emph{static analysis} and \emph{execution}.
During the static analysis phase, GAPS builds a partial call graph, which is then used to construct a sequence of app interaction instructions. Subsequently, GAPS executes these instructions during the execution phase and generates an execution report.

\subsection{Static Analysis Phase}
\label{sec:static-analysis}

The static analysis phase reconstructs execution paths from the app entry points to the target method. For each path, it produces the ordered interaction instructions needed to reach that method. This phase consists of four steps. First, GAPS resolves app entry points by inspecting the manifest and dynamically registered \gls{ICC} components (Section~\ref{design:icc}). Second, starting from the target method, it constructs a partial call graph through a demand-driven backward traversal toward these entry points (Section~\ref{design:cg}). Third, it resolves the branch conditions along each path, retains only satisfiable paths, and records any additional step needed to establish the required program state (Section~\ref{design:conditional_statements}). Finally, it maps the code-level widgets on each path to concrete UI identifiers, producing the actionable GUI elements that the executor triggers at runtime (Section~\ref{design:graphids}).

\subsubsection{Mapping Inter-Component Communication}
\label{design:icc}

GAPS maps ICC by observing that sending an Intent object requires passing it as an argument to a specific Android API call.
The goal is to extract information about both the sender and the receiver of each ICC message. With this information, exported components can serve as app entry points during dynamic interaction, either explicitly (via a fully qualified class name) or implicitly (through an \texttt{action} field).

As shown in Algorithm~\ref{alg:icc}, GAPS first retrieves the ICC information associated with the app components.

\begin{algorithm}[!ht]
\footnotesize
\caption{ICC Mapping}
\label{alg:icc}
\SetKwInput{Input}{Input}
\SetKwInput{Output}{Output}
\Input{
    \textit{components}: Declared components from the manifest and code
}
\Output{
    \textit{icc}: Mapping of components to intent actions, class names, and paths
}
\DontPrintSemicolon

icc $\leftarrow$ empty dictionary \;

\ForEach{$comp \in components$}{
    \If{$is\_exported(comp)$}{
        icc[comp] $\leftarrow$ comp.name \;
    }
    \If{$has\_intent\_filter(comp)$}{
        icc[comp] $\leftarrow$ comp.intent\_actions \;
    }
}

dynamic\_receivers $\leftarrow$ find\_dynamic\_receivers() \;

\ForEach{$recv \in dynamic\_receivers$}{
    reg\_paths $\leftarrow$ get\_registration\_paths(recv) \;
    info $\leftarrow$ data\_flow\_analyze(recv, reg\_paths) \;
    \If{$info$ contains action}{
        icc[recv.class] $\leftarrow$ info.action \;
    }
}

icc\_paths $\leftarrow$ find\_icc\_related\_paths() \;

\ForEach{$path \in icc\_paths$}{
    intent $\leftarrow$ analyze\_intent(path) \;
    target $\leftarrow$ data\_flow\_resolve(intent) \;
    \If{$target$ is not null}{
        icc[target] $\leftarrow$ path \;
    }
}

\KwRet{$icc$}

\end{algorithm}

GAPS begins by inspecting the input components to determine whether they are exported or associated with intent filters (2-9), thereby identifying valid app entry points beyond the main activity. It then analyzes dynamically registered \texttt{BroadcastReceivers} (10), which can be created programmatically and therefore require \emph{data-flow} analysis for mapping. The function \texttt{find\_dynamic\_receivers} (10-12) first locates the code sections that register these components and then applies \textit{data-flow} analysis (13) to extract the relevant metadata, namely the class names and associated \texttt{IntentFilter} objects. An \texttt{IntentFilter} defines invocation parameters such as the \texttt{action}, which is critical for triggering the corresponding receiver (15).
GAPS uses this information to populate the \texttt{icc} dictionary, which maps destination components to their respective \texttt{action} values or fully qualified class names. To comprehensively model all dynamically reachable components, GAPS also analyzes paths involving \gls{ICC} (19-25). This analysis inspects the construction of \texttt{Intent} objects (20) and applies data-flow analysis to resolve their targets (21). The resolved paths are then incorporated into the same \texttt{icc} data structure (23), enabling GAPS to link ICC operations to execution flows.


\subsubsection{Backward Generation of the Call Graph}
\label{design:cg}

GAPS builds call graphs using a target-directed, demand-driven approach that differs from traditional whole-program analyses, such as those used by FlowDroid. Rather than constructing a complete call graph up front, GAPS performs a backward traversal from the target method. This traversal is both inter-procedural and context-sensitive, and it collects only the method calls that lie on feasible paths to the target. The result is a partial call graph slice that is sufficient for reachability while avoiding the scalability bottlenecks of whole-app analysis~\cite{mudflow}. GAPS also incorporates \emph{data-flow} analysis to capture branch conditions and control dependencies, which are essential for reconstructing realistic execution paths, actionable entry points, and GUI event identifiers.
Algorithm~\ref{alg:3} presents the pseudocode for partial call graph generation.

The process starts from the target method and iteratively traverses backward through its potential callers. GAPS maintains a queue of nodes to process (lines 2 and 3), removes the current node from the queue (line 6), and identifies all statically discoverable callers through \texttt{find\_new\_nodes} (line 10). To identify these callers and properly handle object-oriented polymorphism, GAPS constructs a \gls{CHA} graph on demand, iteratively building it by querying the relevant class relationships. This ensures that dynamically dispatched virtual method invocations are conservatively resolved and mapped to all potential caller implementations.
The new caller nodes are added to the call graph and appended to the queue (line 11). When GAPS encounters an entry point, the backward traversal terminates along that branch, and no further predecessor nodes are added. Building on the ICC mapping detailed in Section~\ref{design:icc}, GAPS recognizes an entry point as any framework method that belongs to a component externally reachable through an Intent. This includes life cycle methods invoked by components explicitly exported in the manifest (e.g., \texttt{MainActivity.onCreate}) and dynamically registered ICC components, such as broadcast receivers. Processed nodes are added to the \texttt{seen\_nodes} set (line 12) to avoid cyclic reprocessing and ensure termination (line 7).
This backward expansion continues until no new nodes are found. At that point, the constructed graph contains a precise, target-oriented slice of the app's call graph. Once this partial call graph is complete, GAPS invokes a depth-first traversal (\texttt{dfs\_visit}, line 14) to extract feasible paths from the discovered app entry points to the target method. Finally, GAPS enriches these abstract paths by resolving conditional branches (lines 16 and 17) and extracting the GUI element identifiers (line 18) needed to dynamically trigger the required execution chains.

\begin{algorithm}[!ht]
\footnotesize
\caption{Partial call graph generation}\label{alg:3}
\SetKwInput{Input}{Input}
\SetKwInput{Output}{Output}
\Input{\textit{target\_method}: Target method}
\Output{\textit{result\_paths}: Set of paths inter-procedurally reconstructed}
\DontPrintSemicolon

result\_paths $\leftarrow$ set()\;
nodes\_queue $\leftarrow$ queue()\;
nodes\_queue.add(target\_method)\;
seen\_nodes $\leftarrow$ set()\;
cg $\leftarrow$ graph()\;

\ForEach{$current\_node \in nodes\_queue$}{
    \If{$current\_node \in seen\_nodes$}{
        \textbf{continue}\;
    }
    new\_nodes $\leftarrow$ find\_new\_nodes($current\_node$)\;
    current\_node $\leftarrow$ add\_new\_nodes($new\_nodes$, $cg$, $nodes\_queue$)\;
    seen\_nodes.add($current\_node$)\;
}

result\_paths $\leftarrow$ dfs\_visit($cg$, $target\_method$)\;

\ForEach{$path \in result\_paths$}{
    conditional\_paths $\leftarrow$ find\_conditional\_paths($path$)\;
    path $\leftarrow$ merge\_paths($path$, $conditional\_paths$)\;
    path $\leftarrow$ find\_GUI\_elements($path$)\;
}

\KwRet{$result\_paths$}
\end{algorithm}


\subsubsection{Linking paths that satisfy conditional statements}
\label{design:conditional_statements}
During path reconstruction, as discussed in Section~\ref{design:cg}, GAPS analyzes conditional statements and retrieves the paths that satisfy them.
GAPS currently supports conditional statements involving (i) variables and primitive types (e.g., int, String, float), (ii) objects, which can be compared against \texttt{null}, and (iii) methods, whose return values are considered.
To retrieve the relevant program entities and the constant values they may assume, GAPS combines \emph{points-to analysis} and \emph{constant propagation}.
More specifically, \emph{points-to analysis} determines the set of memory locations (objects or variables) to which a pointer or reference variable may refer, while \emph{constant propagation} aims to replace variables with their constant values whenever possible.
Algorithm~\ref{alg:conditional-resolution} illustrates how these analyses are used to retrieve paths that satisfy conditional statements.

\begin{algorithm}[!ht]
\footnotesize
\caption{Conditional Statement Resolution}
\label{alg:conditional-resolution}
\SetKwInput{Input}{Input}
\SetKwInput{Output}{Output}
\Input{
    \textit{cond}: The conditional statement to analyze \\
    \textit{path}: The execution path containing the conditional statement
}
\Output{\textit{branch\_paths}: List of paths annotated with the conditional outcome}
\DontPrintSemicolon

lhs, rhs $\leftarrow$ extract\_operands(cond)\;

lhs\_points $\leftarrow$ points\_to\_analysis(lhs, path)\;
rhs\_points $\leftarrow$ points\_to\_analysis(rhs, path)\;

lhs\_vals, lhs\_paths $\leftarrow$ constant\_propagation(lhs\_points)\;
rhs\_vals, rhs\_paths $\leftarrow$ constant\_propagation(rhs\_points)\;

branch\_paths $\leftarrow$ empty list\;

\ForEach{$val\_l, p\_l \in (lhs\_vals, lhs\_paths)$}{
    \ForEach{$val\_r, p\_r \in (rhs\_vals, rhs\_paths)$}{
        \If{$verify\_condition(val\_l, val\_r, cond)$}{
            merged\_path $\leftarrow$ merge\_paths(p\_l, p\_r, path)\;
            branch\_paths.add(merged\_path)\;
        }
    }
}
\KwRet{$branch\_paths$}
\end{algorithm}

The algorithm takes as input the conditional statement and the path that contains it.
It begins by extracting the left and right operands (\emph{lhs} and \emph{rhs}, respectively) from the conditional expression (1). For each operand, it performs points-to analysis to trace the origin of the variable or reference in the program (2 and 3). These origin points are then used for constant propagation, which computes all possible constant values the operands may hold, along with the paths where these values are assigned (4 and 5).
After obtaining the constant values and their corresponding assignment paths, the algorithm performs pairwise comparisons for all value combinations (7-13). For each combination, it merges the two associated assignment paths with the current path context and verifies whether the condition could evaluate to true (9). If the condition is satisfiable under that value pair, the merged path is added to the output list of feasible, conditionally valid paths (10-11).
The resulting \emph{branch\_paths} captures all statically inferred control-flow paths that satisfy the conditional statement.


\subsubsection{GUI Event Retrieval}
\label{design:graphids}

To support guided dynamic execution, GAPS identifies the specific GUI elements that trigger event handlers within the app. This enables the generation of interaction plans that accurately simulate user actions. GAPS retrieves these GUI elements statically through a layered application of points-to analysis. Algorithm~\ref{alg:gui-event-resolution} illustrates the procedure.

The process begins by detecting GUI event handlers, such as \texttt{onClick()} or \texttt{onItemSelected()}, that occur within a recognized \emph{activity}. GAPS scans paths for listener registrations, including explicit method calls such as \texttt{setOnClickListener()}. Once a handler is found, points-to analysis resolves the Android object instance linked to the callback (2), such as a \texttt{Button}, \texttt{MenuItem}, or other view element.
After identifying the GUI object, GAPS retrieves the paths in which it appears (3) and applies another round of points-to analysis to trace its instantiation back to common methods, such as \texttt{findViewById()}. From this trace, GAPS extracts the associated resource identifier (e.g., \texttt{button\_main}) by analyzing the arguments passed to the call (5).
The final result is a list mapping each activity to the view identifier corresponding to a GUI event. These events are later embedded into high-level interaction instructions, allowing the dynamic executor to match statically derived GUI elements with actual runtime views.

\begin{algorithm}[!ht]
\footnotesize
\caption{GUI Event Resolution}
\label{alg:gui-event-resolution}
\SetKwInput{Input}{Input}
\SetKwInput{Output}{Output}
\Input{
    \textit{handler}: GUI event handler method \\
    \textit{activity}: Activity where GUI event occurs \\
    \textit{path}: Execution path containing the event registration
}
\Output{\textit{gui\_bindings}: Mapping from activity to resource ID}
\DontPrintSemicolon

gui\_bindings $\leftarrow$ empty list\;

gui\_obj $\leftarrow$ points\_to\_analysis(handler, path)\;

gui\_obj\_paths  $\leftarrow$ find\_paths(gui\_obj) \;

\ForEach{$path\in gui\_obj\_paths$}{
    gui\_id $\leftarrow$ points\_to\_analysis(path)\;
    gui\_bindings.append(activity, gui\_id)\;
}

\KwRet{$gui\_bindings$}
\end{algorithm}


\subsection{Execution Phase}
GAPS' execution phase installs the app on a device and replays the interaction instructions produced by the static analysis, simulating user actions to navigate the UI and trigger the required components. As the app runs, GAPS monitors its execution flow to detect when the target method is reached. When a deterministic step fails because of an unexpected UI state, GAPS falls back to PHIL, an agentic reasoning module that synthesizes context-aware inputs and dismisses unpredictable elements (e.g., system overlays, advertisements, or permission dialogs) to bridge the gap between the static plan and the live interface. 

\subsubsection{Automated Interaction}
\label{design:automated_interaction}

Algorithm~\ref{alg:dynamic_interaction} details the dynamic execution phase of GAPS, which traverses the statically reconstructed paths to trigger the target method at runtime.

The process begins by establishing a runtime monitor (1) on the app to asynchronously detect invocations of the target method. GAPS then systematically evaluates each path in the set of statically reconstructed paths (2). To ensure a clean execution state, the app is restarted at the beginning of every path evaluation (3). GAPS tracks its progress using an instruction index i and initializes an empty map (4 and 5), which records the app screens (activities) where the semantic fallback agent has already intervened.
During the traversal loop, GAPS retrieves the current instruction and the currently displayed activity (7 and 8), then attempts to execute the interaction (9), which may involve interacting with a UI element or sending an intent. If the action is successful, the framework simply advances to the next instruction in the path (10 and 11). However, if the deterministic action fails (e.g., due to a pop-up advertisement or a permission dialog), GAPS invokes PHIL, the fallback agent (14).

\subsubsection{PHIL -- Agentic Fallback}

This agentic intervention is strictly bounded. GAPS checks to ensure PHIL is invoked at most once per activity along a given path (13). If PHIL successfully clears the roadblock, the deterministic loop will naturally retry the stalled instruction on the next iteration. If the roadblock persists or PHIL has already been exhausted for the current activity, GAPS abandons the current path and proceeds to the next candidate path, preventing the system from falling into unbounded exploration loops (18).
Throughout the entire execution cycle, GAPS continuously polls the runtime monitor to check if the target method is executed at any point (21).
Finally, an execution report is returned with measurements on the number of target methods reached.

\begin{algorithm}[!ht]
\footnotesize
\caption{GAPS Dynamic Interaction}
\label{alg:dynamic_interaction}
\SetKwInput{Input}{Input}
\SetKwInput{Output}{Output}
\Input{
    $target\_method$: The target method to reach \\
    $paths$: Set of statically reconstructed paths for $target\_method$ \\
    $app$: Application under test
}
\Output{Execution status (\text{REACHED} or \text{FAILED})}
\DontPrintSemicolon

$\text{install\_monitor}(app, target\_method)$

\ForEach{$path \in paths$}{
    $\text{restart\_app}(app)$\;
    $i \leftarrow 0$\;
    $llm\_used \leftarrow \text{Empty Map}$\;
    
    \While{$i < \text{length}(path)$}{
        $instruction \leftarrow path[i]$\;
        $curr\_activity \leftarrow \text{get\_activity}()$\;
        
        $status \leftarrow \text{perform\_action}(app, instruction)$\;
        
        \eIf{$status == \text{SUCCESS}$}{
            $i \leftarrow i + 1$
        }{
            \If{\textbf{not} $llm\_used[curr\_activity]$}{
                $\text{invoke\_PHIL}(app, instruction, path)$\;
                $llm\_used[curr\_activity] \leftarrow \text{True}$\;
            } \Else {
                \textbf{break} \tcp*{Path failed, proceed to next path}
            }
        }
        
        \If{$\text{is\_method\_reached}()$}{
            \KwRet{\text{REACHED}}\;
        }
    }
}
\KwRet{\text{FAILED}}\;
\end{algorithm}
\section{Implementation}
\label{implementation}

This section describes the implementation of GAPS' two phases. The static analysis phase builds on Androguard~\cite{androguard} and networkx~\cite{networkx} for call graph construction and path reconstruction, and uses Apktool~\cite{apktool} to resolve GUI resource identifiers. The execution phase drives app interaction via AndroidViewClient~\cite{androidviewclient}, with PHIL serving as a fallback AI agent in the event of unexpected UI states.

\subsection{Static Analysis Phase}
\label{impl:cg}
At its core, GAPS uses the \textit{Androguard}~\cite{androguard} Python library to load and analyze input APK or DEX files through the built-in \texttt{AnalyzeAPK} and \texttt{AnalyzeDex}~\cite{analyzeapk} functions.
GAPS then analyzes the \texttt{MethodAnalysis} construct for each method to access its basic blocks.
Each basic block is represented as a sequence of \emph{smali} instructions, which form an intermediate representation of the app's bytecode.
Internally, GAPS uses the \emph{networkx} Python library~\cite{networkx} to construct and model the graph.
GAPS then uses the \texttt{all\_shortest\_paths}~\cite{all_shortest_paths} method from the \emph{networkx} library to retrieve the paths connecting the source nodes and the entry points.
Because solving implicit flows remains an open research problem~\cite{cg_soundness} and falls outside the scope of this paper, GAPS incorporates callback mappings from established Android static analyzers, such as EdgeMiner~\cite{edgeminer} and Soot~\cite{soot}, via its virtual edges~\cite{virtualedges}.
These mappings allow GAPS to find the corresponding constructor or method initializer and partially address implicit flows.
Data-flow analysis is performed over the \emph{smali} code representation, whose rigorous register-based syntax we model to support \emph{points-to analysis} and \emph{constant propagation}.
This \emph{data-flow-aware analysis} is also essential for linking paths that satisfy conditional statements.
Moreover, through \emph{constant propagation}, a conditional statement involving primitive-type variables, objects, or method return values is translated into a comparison over constant values, allowing GAPS to test its satisfiability.

At the end, GAPS uses \textit{Apktool}~\cite{apktool} to retrieve GUI resource IDs.
To convert them from the \emph{smali} numerical representation, GAPS consults the \textit{public.xml} file, where they can be converted into literal identifiers.
GAPS can generate high-level instructions in JSON files that contain actionable entry points, sequences of graphical element IDs to interact with, and the Activity name where they reside. The files also include a call sequence.

\subsection{Execution Phase}
\label{impl:auto_int}
The high-level instructions generated through static analysis allow GAPS to dynamically traverse paths by (i) using one of the app's entry points and (ii) interacting with the GUI. More specifically, the GAPS dynamic module relies on the AndroidViewClient Python library~\cite{androidviewclient} to locate graphical elements via the \texttt{findViewById}~\cite{avc1} and \texttt{touch}~\cite{avc2} APIs, using their literal identifiers, and then interact with them.
Instrumentation tools such as AndroLog~\cite{samhi2024androlog} and ACVTool~\cite{acvtool} can be used in conjunction with GAPS to confirm the execution of target methods. For example, we used AndroLog when evaluating GAPS on the AndroTest benchmark~\cite{androtest} in Section~\ref{sec:evaluation}.
Additionally, Frida~\cite{frida}, a dynamic instrumentation framework, can apply a \emph{hook} to the target method and report when it is executed.

This deterministic approach has limitations for apps that display unpredictable elements in screen overlays. To address unpredictable UI components, we developed and integrated an AI agent, called PHIL, into GAPS' automated interaction module. When GAPS cannot locate an activity or widget on the screen, PHIL guides the interaction with the app until both the widget and the activity appear, enabling GAPS to resume automated interaction under path guidance.
PHIL uses GPT-5.4 as the underlying LLM, with a temperature of 0.7, to flexibly discover alternative paths when resolving complex layout impediments. When invoked, PHIL provides the model with a filtered textual presentation of the active GUI hierarchy, the target method, and the statically retrieved path. The model is asked to return a structured JSON payload that defines an action, such as a ``click'' or ``type'' for text input, and a one-sentence justification for the action. This design choice is motivated by apps that require semantic planning, such as \emph{Yahtzee}, where the number of rounds and players must be specified in the very first activity.
We include PHIL as an optional component that can be disabled or modified to support more LLMs.
\section{Evaluation}
\label{sec:evaluation}

Our evaluation of GAPS revolves around the following three research questions:
\begin{itemize}
    \item \textbf{RQ1}: How does GAPS perform in static path reconstruction compared to static baselines?
    \item \textbf{RQ2}: How does GAPS perform in app targeted execution compared to dynamic baselines?
    \item \textbf{RQ3}: Is GAPS scalable to real-world apps?
\end{itemize}

\subsection{Datasets \& Setup}
\label{datasets}

For the first two research questions (RQ1 and RQ2), we used \androtestApps\ apps from the AndroTest benchmark. We instrumented these apps with AndroLog~\cite{samhi2024androlog} to measure method reachability by logging each method execution. For each app, we randomly selected a fixed set of 50 target methods and reused that set across all tools and runs. This design lets us compare tools on the same reachability problem rather than on different target samples.
The selected targets are not limited to GUI callbacks: only 34.39\% are located within an Activity, meaning that most targets lie elsewhere in the app and are therefore harder to reach through pure GUI exploration. Overall, these statistics indicate that the sample is not dominated by shallow methods near app entry points.

For the final research question (RQ3), we selected the 50 most-downloaded apps from the Google Play Store and targeted the potentially vulnerable methods reported by SPECK~\cite{speck}, a rule-based static analysis tool designed to identify violations of Google's security and privacy guidelines. In this setting, only 5.55\% of the target methods reside in an Activity.

Static analysis experiments are performed on a CloudLab~\cite{Duplyakin+:ATC19} Ubuntu 22.04 machine with 64 GB of RAM. Dynamic experiments are performed on an x86-64 emulator running Android 16, and on a Pixel 2 running Android 11 when an ARM architecture is required.


\subsection{Static Path Reconstruction}
\label{rq1}


RQ1 evaluates the \emph{static} side of target method reachability: given an app and a target method, can we reconstruct at least one path that reaches that method? We focus on two quantities: the fraction of targets for which a path is statically produced and the average analysis time per app. We therefore compare GAPS against the two most relevant baselines for this setting: FlowDroid~\cite{flowdroid}, a strong and actively maintained call-graph generator, and DroidReach~\cite{droidreach}, the only available tool explicitly designed for static path reconstruction.
To ensure a fair comparison, we configured each tool with the strongest method-level setup available to us. FlowDroid was run in its Class Hierarchy Analysis (CHA) configuration, following a recent comparison study~\cite{cg_soundness} that showed this setting yields the best results among the available options. DroidReach reconstructs paths from call graphs generated by AndroGuard~\cite{androguard_cg} or FlowDroid. In our setup, we improved DroidReach by restricting entry points to exported components, whereas the original implementation treats all manifest-declared components as potential entry points. We excluded DroidReach's FlowDroid backend because it relies on an outdated version of FlowDroid, which would have introduced a configuration bias unrelated to the research question. We also considered GoalExplorer for the static comparison, but its static output is limited to an STG that flags target activities rather than reporting how many target methods were actually found; moreover, the current release has broken Maven dependencies, preventing us from extending it to emit comparable statistics.

All experiments used the same fixed set of 50 target methods per app, ensuring that the static tools were evaluated on identical targets. A target was counted as statically reached if the tool produced at least one path containing that method. As discussed in the dataset description, the target sample is not concentrated on near-entry-point methods: only 34.39\% of the targets lie inside Activities, while the remainder are distributed across the rest of the app. Consequently, the experiment tests whether existing static analyses can recover paths to a diverse set of methods, rather than only to shallow callbacks.

The results, summarized in Table~\ref{tab:rq1_static}, show that existing whole-program analysis tools struggle with this task. FlowDroid generated paths for \flowdroidmr\ of the target methods, with an average analysis time of \flowdroidat\ per app. DroidReach performed worse, reaching only \droidreachmr\ in \droidreachat. In contrast, GAPS generated paths for \methods\ of the target methods in just \at\ per app on average. Thus, GAPS is 63.86\% faster than FlowDroid and 45.99\% faster than DroidReach, while uncovering substantially more targets. To characterize the paths reconstructed by GAPS, we analyzed their call-chain lengths (i.e., the number of methods traversed). Across \androtestApps\ applications and 2,609 retrieved sequences, GAPS produced chains ranging from 1 to 39 calls, with an average of 4.52 calls and a median of 2; 17 applications have an average target depth of five or more. By comparison, the paths generated by the baselines are notably shallower. For FlowDroid, we extracted paths from its generated call graphs using NetworkX's \texttt{all\_shortest\_paths} algorithm, yielding an average chain length of 3.35 (median 3.02) and a maximum sequence length of 19 calls. DroidReach produced even shorter sequences, averaging 1.27 calls (median 1.04) and a maximum of 12.

\begin{table}[htbp]
    \centering
    \caption{Comparison of static analysis tools. Reachability indicates the percentage of target methods for which at least one path was generated. Time represents the average analysis duration per application.}
    \label{tab:rq1_static}
    \begin{tabular}{lcc}
        \toprule
        \textbf{Tool}          & \textbf{Reachability (\%)} & \textbf{Avg. Analysis Time} \\
        \midrule
        DroidReach             & \droidreachmr              & \droidreachat               \\
        FlowDroid              & \flowdroidmr               & \flowdroidat                \\
        \midrule
        \textbf{GAPS (Static)} & \textbf{\methods}          & \textbf{\at}                \\
        \bottomrule
    \end{tabular}
\end{table}

These performance and reachability gains stem from architectural differences. Both FlowDroid and DroidReach rely on whole-program abstractions that struggle to provide concrete, executable paths. FlowDroid uses a synthetic \texttt{dummyMain} method, which provides little practical insight into how a path is triggered, whereas DroidReach's entry-point assumptions often miss implicit or programmatically registered triggers. Moreover, neither baseline resolves conditional logic, leaving paths dependent on unchecked branch constraints. GAPS succeeds because its static phase is demand-driven and target-oriented: instead of building and searching a massive whole-program abstraction, it reconstructs only the relevant slice, resolves path conditions, and identifies actionable GUI triggers and entry points.

\begin{tcolorbox}[colback=gray!5!white, colframe=gray!50!black, title=Static Path Reconstruction]
    Static analysis tools fail to build paths to a substantial portion of the target methods: FlowDroid's generated paths reach \textbf{\flowdroidmr} of the methods, while DroidReach reaches only \textbf{\droidreachmr}. In contrast, GAPS generates paths to \textbf{\methods} of the target methods while also being 63.86\% and 45.99\% faster than FlowDroid and DroidReach, respectively.
\end{tcolorbox}


\subsection{Targeted App Execution}
\label{rq2}

RQ2 evaluates whether current GUI interaction tools can solve the \emph{dynamic} side of target method reachability. We therefore consider three representative baselines from different families: APE~\cite{ape}, a strong coverage-oriented GUI tester; Guardian~\cite{guardian}, an LLM-based UI agent; and GoalExplorer~\cite{goalexplorer}, a hybrid activity-guided explorer built on Stoat~\cite{stoat}. This selection allows us to compare pseudo-random/model-based exploration, task-oriented semantic interaction, and screen-transition-guided exploration against GAPS' path guidance under the same reachability objective.

\begin{figure}[!ht]
    \centering
    \includegraphics[width=.95\linewidth]{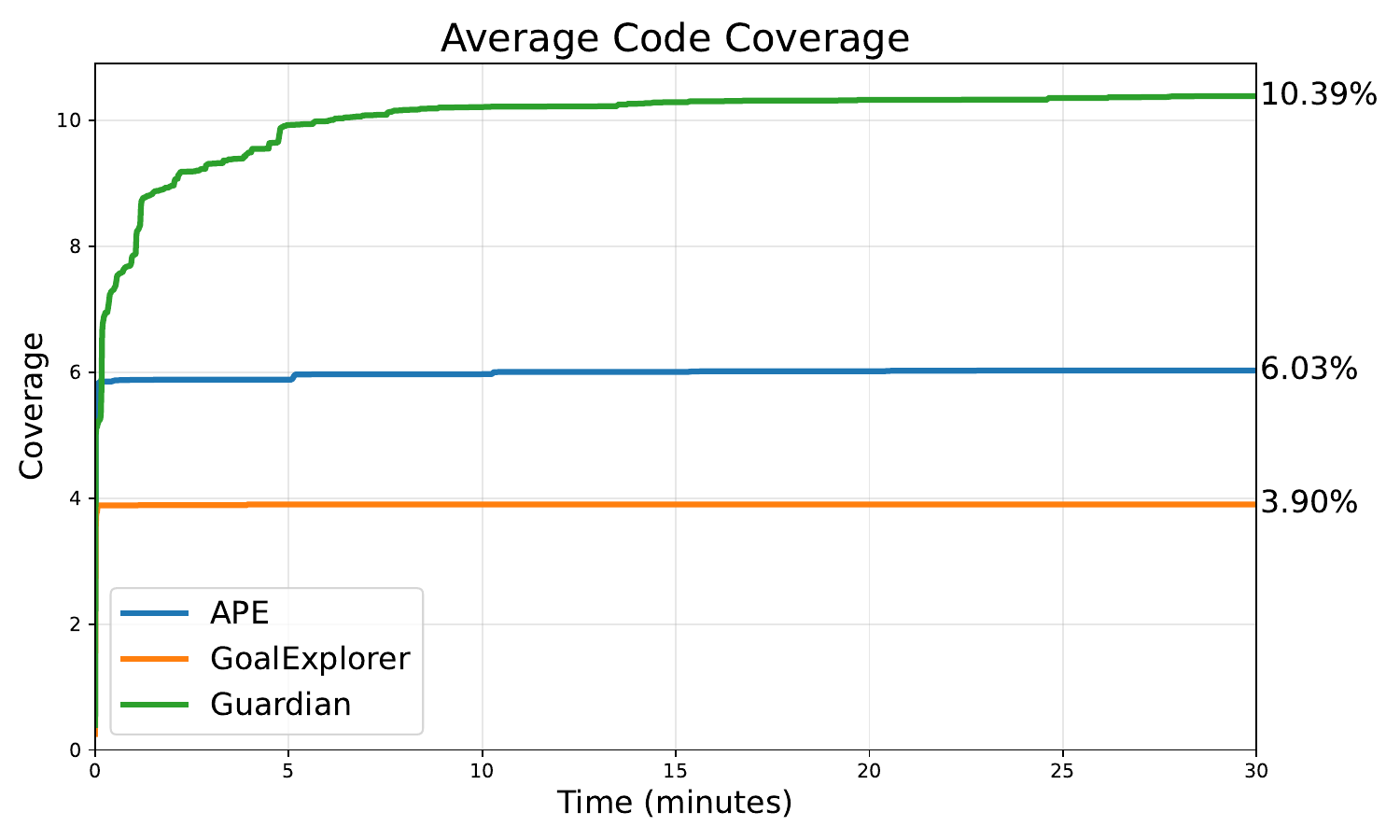}
    \caption{Average coverage over time for the three dynamic baselines used in RQ2, measured three times each with a timeout of 30 minutes.}
    \label{fig:rq2_timeout}
\end{figure}

The experiment was conducted on the \androtestApps\ AndroTest apps. Each app was executed in three independent runs, with a 30-minute timeout per run. We use a uniform timeout for all tools to avoid conflating effectiveness with unbounded exploration time. This is particularly important for tools with different interaction rates and internal overheads. To verify that this timeout does not truncate exploration prematurely, we also examined the average coverage curves of the three dynamic baselines, shown in \autoref{fig:rq2_timeout}. The curves plateau in every case: GoalExplorer and APE reach their maximum coverage within the first few minutes, while Guardian increases coverage more gradually. Since these tools rely on unguided exploration to discover new states, this near-saturation suggests that longer runs are unlikely to materially change the main conclusion of RQ2, even if they slightly increase absolute reachability. 

Guardian requires special methodological care because it was originally designed for user-task-driven UI automation. In RQ2, however, we are not evaluating whether an LLM can be manually coached toward a specific method name; rather, we are evaluating whether an off-the-shelf dynamic interaction tool can reach target methods \emph{without} privileged static guidance. For this reason, we supply the agent with the target classes and method names. We then rely on the LLM's semantic reasoning capabilities to infer the underlying purpose of these targets directly from their identifiers. The model analyzes the current screen state and dynamically prioritizes interactions with UI elements whose representations logically correlate with the inferred behavior of the target methods.
The full prompt is available in our repository. As driving LLM, we used GPT-5.4.

During execution, we collected logs to verify whether each target method was actually executed at runtime. The same target set was reused across all tools, ensuring that differences in reachability are due to the tools themselves rather than to target selection. Although the targets are randomly sampled, the sample is not biased toward easy GUI-local behavior: only 34.39\% of the selected targets are in Activities, while the remaining majority lies outside Activities. As a result, the experiment intentionally includes methods that are less likely to be exposed through shallow exploration, such as targets reached through helper classes, callbacks, and library code.
As detailed in Table~\ref{tab:rq2_dynamic}, across the three runs and the 50 target methods selected for each app, GoalExplorer executed an average of \goalexplorerrr\ methods, while APE reached \aperr\ and Guardian reached \guardianrr.
By contrast, GAPS executed \por\ of the target methods, with an average completion time below 5 minutes.

\begin{table}[]
    \centering
    \caption{Comparison of dynamic execution tools over three runs with a 30-minute budget per run. Reachability indicates the percentage of target methods successfully executed at runtime.}
    \label{tab:rq2_dynamic}
    \begin{tabular}{lcc}
        \toprule
        \textbf{Tool}            & \textbf{Reachability (\%)} & \textbf{Avg. Runtime} \\
        \midrule
        GoalExplorer             & \goalexplorerrr            & 00:30:00 (Timeout)    \\
        APE                      & \aperr                     & 00:30:00 (Timeout)    \\
        Guardian                 & \guardianrr                & 00:30:00 (Timeout)    \\
        \midrule
        \textbf{GAPS (w/o PHIL)} & \porWoPhil                 & \textbf{00:02:26}     \\
        \textbf{GAPS (Dynamic)}  & \textbf{\por}              & 00:03:15              \\
        \bottomrule
    \end{tabular}
\end{table}

These results highlight complementary limitations of current GUI interaction tools for target method reachability. APE can rapidly generate events, but its coverage-oriented strategy often struggles to construct the specific app state required to trigger deeper behaviors. GoalExplorer benefits from the STG, yet its guidance is centered on activities and screen transitions; consequently, it is less effective when the target method resides in a third-party library or when execution depends on control-flow conditions that are not resolved by reaching the right screen. Guardian can reason semantically about visible UI elements, but semantic task planning alone remains insufficient when target execution depends on hidden callbacks, non-UI components, or code-level conditions that are not visible in the current interface.
The lower average runtime of GAPS further reflects that path-guided execution often reaches the target before the timeout, rather than exhausting the entire exploration budget.
Compared with APE and Guardian, GAPS avoids unguided exploration by transforming static reachability information into a concrete execution plan. Compared with GoalExplorer, whose guidance remains centered on screens, activities, and STG transitions, GAPS reasons directly about method-level targets and can therefore better handle targets reached through callbacks or guarded paths.
The gap between GAPS's static and dynamic performance also clarifies its remaining failure cases. Although GAPS can statically generate paths to \methods\ of targets, dynamic execution reaches only \por\ of them because some statically recovered paths still require hard-to-synthesize runtime states, dynamically generated interfaces, or external inputs and intents. This interpretation is consistent with the failure modes discussed in Section~\ref{discuss}.

To further quantify the impact of PHIL, we conducted an ablation study that evaluates GAPS without the LLM-based module. When relying strictly on deterministic execution of the static plan, without any agentic intervention, GAPS's dynamic reachability drops from \por\ to \porWoPhil. This substantial decrease highlights the importance of semantic resolution in modern app testing. Contemporary Android apps rely heavily on dynamically rendered content, such as advertisements and runtime permission dialogs, which can obscure expected GUI elements and disrupt deterministic traversal. Without PHIL to synthesize context-aware inputs and dismiss these unpredictable roadblocks, the static blueprints frequently stall. PHIL therefore plays a vital role in pursuing valid execution paths that would otherwise fail, bridging the gap between static analysis and the stochastic nature of Android apps.

\begin{tcolorbox}[colback=gray!5!white, colframe=gray!50!black, title=RQ2: Targeted App Execution]
    Current GUI interaction tools struggle with target-driven method reachability. Under a strict 30-minute timeout and on the same fixed target set, GoalExplorer reached only \textbf{\goalexplorerrr}, APE \textbf{\aperr}, and Guardian \textbf{\guardianrr} of the target methods. Saturation curves confirm that this gap is not an artifact of the timeout. In contrast, GAPS dynamically executed \textbf{\por} of the targets in under 5 minutes on average. These findings show that unguided exploration, screen-level guidance, and pure LLM-guided task planning are insufficient, motivating GAPS's hybrid integration of path-sensitive static guidance.
\end{tcolorbox}

\subsection{GAPS Applied to Real-World Apps}
\label{rq3}


RQ3 assesses the broader applicability of GAPS to large, real-world apps and security-relevant targets. Unlike the AndroTest experiments, where target methods are randomly sampled, this setting uses the method locations reported by SPECK~\cite{speck} as potentially vulnerable. We collected the 50 most downloaded apps from the Google Play Store in June 2026 and ran GAPS on them; the full list of apps and results is available in our repository. This setting is therefore closer to the intended use case of GAPS: an analyst starts from suspicious code locations and wants to determine whether those methods are reachable at runtime in the analyzed app.

The measurement setup also differs from that of AndroTest. For real-world apps, AndroLog~\cite{samhi2024androlog} failed to repackage APKs to enable method-level logging, so we used GAPS' Frida integration to dynamically hook target methods and record each hook invocation. A target was counted as statically reached if GAPS generated at least one path to it, and as dynamically reached only if the Frida hook confirmed that the method was executed during guided interaction. This makes the dynamic measurement stricter than simple screen completion or UI coverage, since it requires direct evidence that the specific target method was executed.

Under this configuration, GAPS successfully generated paths to \rwaMR\ of the target methods, yielding 219 call sequences in total. These sequences ranged from 1 to 41 methods, with an average of 12.42 calls and a median of 4; 45 apps had an average call sequence length of 5 or more. Static analysis required an average processing time of \rwaAT\ per application. This result shows that the static phase remains practical even on large commercial apps. At the same time, the lower static reachability compared with AndroTest highlights the increased difficulty of real-world apps: they contain more obfuscation, more third-party frameworks and libraries, and more complex callback and inter-component structures, all of which make target-oriented path reconstruction harder.

During dynamic analysis, GAPS successfully executed an average of \rwaRR\ of the target methods over three runs, with an average execution time of 4 minutes and 48 seconds. The gap between \rwaMR\ and \rwaRR\ is informative because it marks the boundary between static feasibility and practical execution. In many cases, GAPS can reconstruct a plausible path but cannot automatically traverse it because commercial apps require a richer runtime state than benchmark apps (e.g., app-specific textual input, dynamic UI rendering, payment or authentication steps, or intents that require permissions or extra parameters). In other cases, the path may pass through heavily obfuscated or framework-driven code that is visible to static analysis but difficult to realize through ordinary user interaction. Thus, the real-world evaluation shows both that GAPS scales to practical, security-relevant targets and that the remaining failures stem largely from runtime state reconstruction rather than from static path synthesis alone.

\begin{tcolorbox}[colback=gray!5!white, colframe=gray!50!black, title=RQ3: GAPS on Real-World Apps]
    GAPS remains practical on real-world, security-relevant targets. On 50 top Google Play Store apps, it generated paths to \textbf{\rwaMR} of the target methods with an average static analysis time of \textbf{\rwaAT}, and achieved Frida-confirmed dynamic execution for \textbf{\rwaRR} of them. The gap between static and dynamic reachability indicates that the main remaining challenge in commercial apps is reconstructing the required runtime state, rather than merely recovering candidate paths.
\end{tcolorbox}

\section{Discussion and Limitations}
\label{discuss}

While GAPS demonstrates improvements in the reachability of targeted methods, it comes with some limitations.

First, its reliance on static bytecode analysis limits its effectiveness for apps built with frameworks such as \emph{Flutter} or \emph{React Native}, where the app's logic resides outside traditional Dalvik code.
Furthermore, apps with an obfuscated call graph can lead to path explosion and increase analysis time, as evidenced by the longer analysis times of real-world apps.
Another obstacle to path reconstruction is \emph{dead code} located in libraries, which can worsen GAPS' performance. Finally, call graph generation is known to suffer from \emph{unsoundness}~\cite{cg_soundness}, caused by untracked implicit flows. GAPS uses the same callback mappings used by FlowDroid~\cite{flowdroid} and Soot~\cite{soot}, and also includes those generated by EdgeMiner~\cite{edgeminer}, but might still suffer from unsoundness when an unexpected implicit flow is met. The analysis of the Android Framework to identify implicit edges for static analysis tools is an orthogonal problem that falls outside the scope of this work.
Our current approach does not support Jetpack Compose. Unlike XML-based layouts, Jetpack Compose defines UI elements directly in Kotlin code and attaches event handlers through lambdas rather than OnClickListener objects. Furthermore, UI state can change dynamically through recomposition. These differences make data-flow analysis for Jetpack Compose more complex. Extending static analysis to reliably extract GUI elements and events from Jetpack Compose remains an open research problem, and we consider this a direction for future work.

In some cases, we cannot dynamically execute target methods, even though GAPS has built a static path. Complex interactions might be required to fully traverse a path: \emph{Hot Death} and \emph{Yahtzee} are games where some of the randomly selected target methods can only be reached upon winning a game. Finally, most apps have paths with intent filters as entry points, which can have permissions or be owned by the system (e.g., \texttt{BOOT\_COMPLETED}), or require data to be passed along with the Intent. Retrieving the required data parameters and automatically generating Intents are orthogonal to dynamic analysis and fall outside the scope of this paper.
Additionally, modern applications frequently embed dynamically rendered layouts or WebViews that hinder static analysis. While GAPS successfully leverages PHIL to navigate these unpredictable UI roadblocks at runtime, the reliance on an LLM inherently introduces a degree of non-determinism. To mitigate this, GAPS strictly confines PHIL to a bounded fallback mechanism, and our evaluation reports the average of three independent runs per app to account for variance.

Despite these challenges, GAPS's hybrid approach strikes a balance between precision and scalability, providing a practical foundation for future enhancements.
\section{Conclusion}
\label{conclusions}

In this paper, we present GAPS, a framework designed to statically reconstruct paths that guide dynamic analysis until the target method is executed. 
In our evaluation, we show that GAPS can reconstruct at least one path for \methods\ of the target methods and traverse \por\ of these paths in apps from the AndroTest benchmark, with an average analysis time of \at.
In contrast, state-of-the-art tools achieve lower performance in static path reconstruction and dynamic method reachability: GoalExplorer (\goalexplorerrr), APE (\aperr), Guardian (\guardianrr), FlowDroid (\flowdroidmr), DroidReach (\droidreachmr), with longer or less targeted analysis phases.
On the other hand, results on the real-world apps dataset show that GAPS can reconstruct paths for \rwaMR\ of the target methods, averaging an analysis time of \rwaAT\ per app and dynamically executing \rwaRR\ of them.

Overall, we conclude that GAPS bridges static and dynamic analysis, guiding app exploration towards selected target methods.

\newpage
\balance

\section{Data Availability}
We release GAPS' source code, its evaluation data, and setup in open-source~\footnote{\url{https://github.com/samudoria/GAPS/}}. Additionally, we have included instructions to reproduce GAPS' results on the AndroTest benchmark.

\bibliographystyle{ACM-Reference-Format}
\bibliography{references}

\end{document}